\documentstyle[12pt,twoside,fleqn,espcrc1]{article}

\newcommand \beq{\begin{equation}}
\newcommand \eeq{\end{equation}}

\def\frac#1#2{{#1 \over #2}}

\def\half{\ifinner {\scriptstyle {1 \over 2}}
   \else {1 \over 2} \fi}

\def\simge{\mathrel{%
   \rlap{\raise 0.511ex \hbox{$>$}}{\lower 0.511ex \hbox{$\sim$}}}}
\def\simle{\mathrel{
   \rlap{\raise 0.511ex \hbox{$<$}}{\lower 0.511ex \hbox{$\sim$}}}}

\def\slashchar#1{\setbox0=\hbox{$#1$}          
   \dimen0=\wd0                             
   \setbox1=\hbox{/} \dimen1=\wd1              
   \ifdim\dimen0>\dimen1                       
      \rlap{\hbox to \dimen0{\hfil/\hfil}}     
      #1                                       
   \else                                       
      \rlap{\hbox to \dimen1{\hfil$#1$\hfil}}  
      /                                        
   \fi}

\def\subrightarrow#1{%                          % #1 under arrow
  \setbox0=\hbox{%                              % set a box
    $\displaystyle\mathop{}%                    % no mathop
    \limits_{#1}$}%                             % just limits
  \dimen0=\wd0%                                 % get width
  \advance \dimen0 by .5em%                     % add a bit
  \mathrel{%                                    % space like =
    \mathop{\hbox to \dimen0{\rightarrowfill}}% % arrow to width
       \limits_{#1}}}                           % text below

\def\journal#1#2#3#4{\ {#1}{\bf #2} ({#3})\  {#4}}

\def\NPA{\journal{Nucl.\ Phys.\ {\bf A}}}
\def\NPB{\journal{Nucl.\ Phys.\ {\bf B}}}

\def\PLB{\journal{Phys.\ Lett.\ {\bf B}}}

\def\PRD{\journal{Phys.\ Rev.\ {\bf D}}}
\def\PRL{\journal{Phys.\ Rev.\ Lett.}}
\def\PhysRept{\journal{Phys.\ Repts.}}

\def\ZPhysC{\journal{Z.\ Phys.\ C}}

\def\picture #1 by #2 (#3){
  \vbox to #2{
    \hrule width #1 height 0pt depth 0pt
    \vfill
    \special{picture #3} % this is the low-level interface
    }
  }

\def\scaledpicture #1 by #2 (#3 scaled #4){{
  \dimen0=#1 \dimen1=#2
  \divide\dimen0 by 1000 \multiply\dimen0 by #4
  \divide\dimen1 by 1000 \multiply\dimen1 by #4
  \picture \dimen0 by \dimen1 (#3 scaled #4)}
  }

\def\tilQ{\mbox{v\cdot q}}
\def\tilQ1{\mbox{$v\cdot q_1$}}
\def\tilQ2{\mbox{$v\cdot q_2$}}

\newcommand{\AmS}{{\protect\the\textfont2
  A\kern-.1667em\lower.5ex\hbox{M}\kern-.125emS}}

\title{Signals of the quark-gluon plasma in nucleus-nucleus collisions}

\author{Jean-Paul Blaizot\thanks{Affiliated to CNRS}\address{CEA-SPhT ,
Orme des merisiers,  \\ 
        91191 Gif-sur-Yvette Cedex, France}%
    \thanks{SPhT is a laboratory from  Direction des Sciences
de
la Mati\`ere du Commissariat \`a l'Energie Atomique}    }

\begin{document}
% typeset front matter
\maketitle

\begin{abstract}
This talk is a brief overview of the present status of our understanding of   
nucleus-nucleus  collisions at high energy and the search for signals of the
quark-gluon plasma. 
\end{abstract}

\section{INTRODUCTION}

The main motivation for studying  nucleus-nucleus collisions at high energy is to
learn the properties of the densest and hottest forms of matter that one can
produce in the laboratory. One hopes in particular to reach the conditions under
which  hadronic matter is expected to turn into the quark-gluon plasma, a new phase of
matter whose degrees of freedom are the hadron constituents, the quarks and
the gluons.  Understanding
the behaviour of bulk matter governed by QCD elementary degrees of freedom and
interactions, and studying how it turns into hadronic matter, offers  challenging
perspectives and touches fundamental issues in the study of  Quantum Chromodynamics in
its non perturbative regime, such as the nature of confinement, of  chiral symmetry
breaking, etc.

Unfortunately, the tools with which  we are probing these fascinating features of
dense and hot matter are  not ideal. The dynamics of nuclear
collisions is complicated and at present allows at
best for only semi-quantitative predictions. In absence of a definite signal to look
for, we need to learn details of  how nucleus-nucleus collisions work before
one can draw any general conclusion about  properties of hadronic matter. Progress in
the field is therefore largely conditioned  by progress in experiments  which, in
fact, has been quite impressive. 

The data accumulated over the last 10 years both  at the BNL/AGS and at CERN/SPS, in
particular those involving the heavier projectiles and targets, start to draw a
consistent picture of what happens in nucleus-nucleus collisions at high energy.
There is clear evidence that at the highest energy achieved so far, nuclear
collisions deviate substantially form a naive picture based on a mere superposition 
of independent nucleon-nucleon collisions;  collective behaviour is seen. As the
collision energy is tuned up the relevant degrees of freedom change, from nucleons to
hadronic  resonances and hadronic strings, and hints that quark degrees of freedom
are playing a role have been obtained.

 But while a  coherent
picture of the collision dynamics  is
emerging, finding unambiguous signatures of quark-gluon plasma
formation remains an open problem. Presumably, unless one is very
lucky, confirmation of plasma production will not come from a unique signal,
and evidences based on
systematic and well focused observations will have to be accumulated. 
Some may wish to argue that, perhaps, we are lucky. Indeed, several anomalies have
been observed in the data.  Among these are 
 the suppression of $J/\Psi$ production, the excess emission of lepton pairs in
the mass range below the
$\rho$ resonance and  the enhanced production of strange  and multistrange  baryons.
The temptation is great to associate these anomalies with the production of the
quark-gluon plasma but, in my opinion, this is  a bit premature.

This Quark-Matter meeting is special. It is the last conference before RHIC
starts to produce data, opening a new area in the field. This is also the time where the CERN SPS program is coming
to an end.  RHIC physics will be reviewed in a special session organized by
M.~Gyulassy, while E.~Shuryak will discuss the CERN SPS program and its
future perspectives. The present talk is a brief introduction to the field. My goal
is to indicate where the focus of the present discussions is without going into
the details of the interpretations of the various results. Recent, more systematic,
reviews can be found in Refs.
\cite{Bass98,Heinz99,Stachel99}, and there is some overlap, unavoidable, with
E.~Shuryak's talk 
\cite{Shuryak99}. 
The talk is organized as follows. I start with theoretical
considerations on the phase diagram of hot and dense
hadronic matter  and the properties of the quark-gluon plasma. Then  I review
the general patterns of ultrarelativistic collisions.   In the last part of the talk
I briefly discuss  the specific signals  for which anomalous
behaviour has been observed.  The last section contains conclusions.

\section{THEORETICAL CONSIDERATIONS}

There is some  rough 
 analogy between the transition from hadronic matter to the quark-gluon plasma
and that  from  a neutral atomic gas to the corresponding ionized   plasma. In both
cases as the temperature or the density rises, the basic degrees of freedom in the
system change, becoming, in the high temperature/high density phase, the elementary
constituents. However, in spite of the fact that the neutral atomic gas and the
completely ionized plasma have very distinct physical properties,  no phase transition
separates them, and the process of ionization  is a very
gradual one.  In contrast, QCD predicts that the transition from hadronic
matter to the quark-gluon plasma is a sharp one, accompanied 
by the rapid increase of the entropy density corresponding to the liberation of quark
and gluon degrees of freedom. Because of QCD asymptotic freedom, it is not surprising
that quarks and gluons become free at high temperature (or high density); what is not
a priori obvious is the sharpness of the transition, and also the fact that it occurs
for a relatively low temperature. The critical temperature for pure gauge theory is
now determined with an  accuracy of a few percent 
\cite{Boyd96}
 in pure SU(3) gauge theory: $T_c\simeq 264$ MeV. With dynamical quarks,
calculations are more complicated and the resulting critical temperature more
uncertain, $T_c\simeq 150-200$ MeV.

The nature  of this (phase) transition has been somewhat clarified, but not
entirely 
\cite{Laermann96}. In  pure SU(3) gauge theory, one has evidence of a  first
order transition. With massless  or light
quarks, the transition seems to be dominated by the effects of chiral symmetry
breaking and the associated soft modes. Universality arguments suggest then an O(4)
critical behaviour for 2 light flavors, whereas for  3 massless quarks  the
transition is first order. What happens in the real world depends on whether the
mass of the strange quark can be considered as large or small. A heavy strange quark
is  inert in the transition which is then, as for two flavors, a
second-order one. If the strange quark is effectively light, the transition is first
order. Unfortunately,  the actual mass of the strange quark is
of the order of the typical QCD scale, and the situation is still 
controversial. It cannot be excluded  that with non vanishing masses for all quarks,
the sharp  phase transition  disappears and becomes simply a  crossover. 

 How can we characterize the phases before and
after the transition?  This question, related to that of finding an unambiguous 
signature
of the quark-gluon plasma, has no simple answer. In the real world where quark masses
are non vanishing,  no order parameter has been found to distinguish the two phases.
What is meant then by  confinement or deconfinement transition? A plausible picture
which is receiving support from lattice calculations is that of the dual
superconductor involving color magnetic monopole condensation. This picture  is
reviewed here  by A. Di Giacomo
\cite{DiGiacomo99}. One interesting consequence of this picture is the existence of
strings between heavy quarks. This leads to a linearly increasing potential at zero
or low temperature, and provides a simple view
of color confinement. Quite remarkably, the
string tension drops rapidly as $T$ approaches
$T_c$ and vanishes at $T_c$
\cite{Laermann95}. Above $T_c$ the potential is a screened potential, with the
screening radius a decreasing function of the temperature. We shall refer to this
behaviour in our discussion of $J/\Psi$ suppression. 

Another perspective on what happens at the transition is given by chiral
symmetry. The quark condensates are expected to decrease with increasing temperature
and density  and, in the case where the quarks are massless, to vanish for some
values of these parameters: at this point chiral symmetry is restored. In the limit of
small temperature
$T$ and baryon density
$\rho$, the  variations of the condensates can be obtained from model independent
considerations: 
$\delta 
\langle\bar q q\rangle_T\propto -{T^2}$ and 
 $\delta \langle\bar q
q\rangle_\rho\propto -\rho.
$
For higher density however,   interaction effects must
be taken into account; these may strongly affect in particular the value of the
density at which chiral symmetry is restored \cite{Weise96}.

Lattice results  show  that the ideal gas
limit is approached as $T$ becomes large, but this approach 
is slow: typically, the energy density at
$2T_c$ is about 85\% of the Stefan-Boltzmann limit value. These results  
can be accounted for reasonably well by
phenomenological fits involving  massive quasi-particles \cite{Peshier}. 
Although the
quasiparticle picture suggested by such fits is a rather crude one, it supports 
the idea that one
should be able to give an accurate description of the
thermodynamics of the quark-gluon plasma in terms of its elementary excitations and
encourages the development of analytical calculations of the thermodynamic potential
using weak coupling techniques. Such calculations are difficult: although the
gauge coupling $g$ is small if the temperature $T$ is sufficiently high, the
perturbative series shows rather  poor convergence properties (see \cite{QCDP}, and
also
\cite{Pade,KPP,DHLR2}). 
 However, sophisticated rearrangements of the perturbative expansion and various 
resummations have been applied to the calculation of the   thermodynamic potential
\cite{ABS}. Particularly promising in this context are  self-consistent
approximations  of the entropy which have been shown recently to accurately 
reproduce lattice data at high temperature 
\cite{BIR99}. Although much remains to be done,  the results  obtained so far are
quite encouraging and give us the hope that an analytical control of the high
temperature phase of QCD is  within reach.

A main limitation of present lattice calculations is their inability to
deal with finite chemical potentials. There is little progress to report on this
issue, although  some of the pathologies of the
quenched approximation have been clarified with the use of random
matrix models \cite{Stephanov96}.
There has been however recently  exciting  developments  in analytical
investigations of the high density part of the phase diagram with the resurgence of
the old idea of color superconductivity\cite{Bailin84}, but with totally new
perspectives. Among all possible quark pairings, it has been recognized in
particular that a special condensate involving a correlation between color and flavor,
possible with  massless flavors, has very special properties. This  ``color-flavor
locked state'' 
\cite{Alford98} leads to
substantial gaps, of the order of 100 MeV and has many remarkable features.  A most
intriguing  one is the apparent continuity between  this phase of
quark matter and  some form of nuclear matter, as it has been suggested in
Ref.~\cite{Schaefer99}. This may offer  the chance to to explore  such properties as
confinement or chiral symmetry breaking using  weak coupling
calculations at high density,  a most fascinating possibility indeed!

I would like to end this section by returning to the real world and list a few {\it
questions for experiments}. Clearly, given the
complexity of nuclear collisions, we  cannot probe all the detailed features
discussed above. So what can we hope to ``see''?
 Can we observe
changes in the equation of state, see
the  phase transition?  
 Can we trace  deconfinement, or  chiral
symmetry restoration,  in an unambiguous way?  
Does  matter produced at the early stages
of the collisions behave  as  non interacting quarks and
gluons? 
  Note that most of these questions refer to systems
in thermal equilibrium; it is for such systems that our theoretical tools
are best developed. Colliding nuclei are not at all systems in equilibrium,
although there is some evidence that local thermal equilibrium may 
be achieved at the
late stages of the collisions. 
There may exist interesting genuine  non equilibrium
phenomena, an  example being provided by the so-called Disoriented Chiral Condensate 
\cite{Rajagopal95}. Finally let  us remark that in all these experimental studies we
have very few  control parameters 
 at our disposal, essentially the nuclear sizes (and the impact parameter) and the
beam energy. However, data with high statistics  allow various cuts  and may
effectively provide new ones.

\section{NUCLEUS-NUCLEUS COLLISIONS. GENERAL PATTERNS}

Measurements of the transverse
energy distributions \cite{Barrette93,Alber95} provide access to the energy
density achieved in the collisions. Simple
estimates using     Bjorken's formula, $$
\epsilon_0={1\over{\tau_0 \pi R^2}}{{\rm d}E_T\over {\rm
d}y},
$$
lead  to  $\epsilon_0\approx 1.3$ GeV/fm$^3$ at the  AGS (${\rm
d}E_T/{\rm d}\eta=200$ GeV for Au+Au central) and $\epsilon_0\approx  3$ GeV/fm$^3$ 
at the SPS (${\rm d}E_T/{\rm d}\eta =450$ GeV at SPS for Pb-Pb), taking for
$\tau_0$ the generic value of 1 fm/c. Although they should be viewed as
crude estimates, these numbers indicate that {\it appropriate conditions are
possibly met} for the formation of a transient quark-gluon plasma at the SPS.   
The measurement of baryon densities reveal a large
baryon stopping at the SPS
\cite{Roland98},  larger than expected,  although the maximum baryon density  is there
lower than that achieved at the AGS.
 
There is  evidence that hadronic matter at freeze-out is nearly
``thermal" (for a review see \cite{Stachel96}). Rather remarkably,  particle
ratios are well fitted by simple statistical models involving only two parameters,
 a  temperature
$T_f$ and a baryon chemical potential
$\mu_B$. One should distinguish here between the {\it chemical freeze-out} at which
point matter composition is frozen, from the {\it thermal freeze-out} where particles
undergo their last collisions.   Particle ratios determine the parameters of the
chemical freeze-out, while the parameters of the thermal freeze-out can be obtained
with additional information from momentum distributions. 
 Transverse momentum spectra are seen to be ``blue-shifted'', reflecting
the collective flow of particles moving towards the observer. This effect is seen on
spectra  as an increase of the inverse slope (effective temperature) with the mass of
the particle considered.
The freeze-out temperature and the collective expansion velocity
thus determined are 
compatible with interferometry measurements \cite{Wiedemann99}.

  Thermal freeze-out occurs
at
 a temperature lower than chemical freeze out, that is, given the expansion, at
a later time. Freeze-out parameters evolve with beam energy, moving from high
density/low temperature at small beam energy to low density/high temperature at high
beam energy. These parameters can be displayed in a phase diagram which has been shown
many times at the last Quark Matter meeting \cite{PBM96a}.  A striking feature
 of this diagram is that the line representing
matter at freeze-out is quite close to that 
 representing the phase boundary toward the quark-gluon plasma \cite{PBM96b}.  
Another interesting observation is that 
 the energy per particle on the freeze-out line 
is about 1 GeV \cite{Cleymans99}.
Although they are highly suggestive, the significance of all these observations is
still unclear. The fact that similar models reproduce the particle production in $e^+
e^-$ collisions
\cite{Becattini96}
 signals  a universal behavior, reminiscent of the Hagedorn's
picture\cite{Hagedorn60},  and suggests that  phase space is statistically populated
in the prehadronization phase.

Various {\it collective flows} have been observed (for a review see
\cite{Ollitrault98}, and Danielewicz at this meeting).  For
 central  collisions, particle emission is azymuthally symmetric, and leads to 
transverse or radial flow responsible for the distortions of the momentum
distributions referred to earlier. In {\it non central collisions},  directed
flow and elliptic flow can occur. The nature of the elliptic flow
changes with the beam energy, from  low energy where it is out of plane (also referred
as squeeze-out), to high energy where the flow is enhanced in the direction of
the impact parameter in which  the pressure gradient is the strongest.  Analysis of
the  various flow patterns, and their evolution with  the collision energy, 
can give information on the equation of state.

 With the increasingly accurate measurements 
 of two particle correlations and interferometry \cite{Bearden96},
the space-time picture of nucleus-nucleus collisions at high energy is
altogether becoming more and more precise.
New perspectives are  offered by analysis of 
{\it event by event } fluctuations
\cite{Appelshauser99,Stephanov99}, which are made possible by the high statistics
data now available.

\section{SPECIFIC SIGNALS}

I turn now to specific ``signals'', that is observables for which anomalous behavior
has been detected. At least for two of them, namely the enhancement of strangeness
\cite{Rafelski82,Koch86} and the suppression of
$J/\Psi$ production \cite{Matsui86}, the observed effects were anticipated, although
it is fair to say that the  predictions were not quantitative: the results obtained
in Pb-Pb collisions came as a surprise, and their interpretation is still under
debate. On each of these subjects there is much to say (strangeness for instance
having  its own topical meetings!) and I shall only give a few indications.

\subsection{Strangeness}

An enhancement of strangeness 
(properly
defined) is observed in all
experiments which measure particles carrying strange quarks: $K$, $\phi$, $\Lambda$,
$\Xi$,
$\Omega$. 
While at the AGS most of strangeness production can be  understood in
terms of independent nucleon-nucleon collisions and hadronic
reinteractions \cite{Odyniec98}, this is not so at the SPS. In particular precise
measurements of the individual hyperon yields ($\Lambda$, $\Xi$,
 $\Omega$, and the corresponding antiparticles) in Pb-Pb collisions 
has revealed a systematic increase with respect to p-Pb as the strangeness content of
the particle increases \cite{Antinori99}. This observation  is difficult to account
for  by invoking hadronic rescattering,  and in fact none of the hadronic models
which successfully reproduce the bulk of the data fit the strange particle yields
without ad hoc adjustments.  Note that
$K$ and
$\Lambda$ are relatively easy to produce in a hadron gas, but  the cross section for
the production of 
$\Omega+\bar\Omega$  is  small \cite{Koch86}.
One could imagine a scenario where multistrange hadrons are  produced
in steps, but this seems to take more time than is available. 

Another important observation is that the mechanism responsible for
strangeness enhancement in Pb+Pb collisions seems to be independent on the centrality
 when the number of participants is 
$N_{part}\simge  100$ \cite{Antinori99}.  It is clearly
important to explore smaller systems or more peripheral collisions to determine
where the effect sets in. A similar remark applies to the energy dependence of the
effect, and pinning down the onset of the phenomenon as the function of beam
energy is one of the most compelling motivations to perform a low energy run at
the CERN/SPS. 

These observations, combined  with the remarks made earlier on the properties of
matter at freeze-out, and in particular the fact that the ratios of strange particles
are compatible with statistical models \cite{Bialas98,PBM99}, convey the impression 
that much of the   strangeness observed in Pb-Pb collisions is already present in the
early stages of the collision. 

\subsection{Dileptons}

The vector mesons, through their decay into dileptons, provide access to
the properties of the dense matter at various stages in the collisions. The $\rho$
meson plays a particular role because its lifetime (1-2 fm/c) is such
that it decays much of the time while being in matter. Since the dileptons produced
in the decay interact weakly with the surrounding matter they 
carry direct information about the state of matter at the time of the
decay. 

The CERES Collaboration \cite{ceres99} has obtained evidence for a significant excess
of dileptons with invariant mass below that of the $\rho$. The  excess, which  is 
concentrated at low transverse momentum 
$p_T$, is not accounted for by $\pi^+ \pi^-$ annihilation
in vacuum. Typical dilepton production processes involve  the $\rho$ meson as an
intermediate state, and the $\rho$ meson can be affected by the
surrounding medium. 

There has been much effort devoted lately to understanding how the 
basic properties of  hadrons are modified in matter. The way 
hadronic interactions modify the spectral density of the $\rho$ meson, leading in
particular to a shift of its mass and an increase of its width,  
has been analyzed with increasingly sophisticated theoretical models (see for
instance \cite{Wambach98} and the talk by R.~Rapp at this meeting). Certainly the
most exciting issue here is the potential relation of  such modifications to the
onset of chiral symmetry restoration,  which could occur under the conditions
realized at the SPS.

Much progress in the field is expected from the coming runs which
should provide high precision data, allowing in particular to distinguish the
behaviors of the various resonances in matter, while the low
energy run should demonstrate the sensitivity of  the effect to a change
in the  the baryonic density.

\subsection{$J/\Psi$ suppression}

With the $J/\Psi$, a tiny bound state made of a pair of  heavy charm  quark  and
antiquark, we can probe the very early stages of the collisions. Contrary to
the
$\rho$, the $J/\Psi$ has a very long lifetime and it decays into dileptons 
only when it is far from the collision zone. However, as pointed out by  Matsui and
Satz \cite{Matsui86},  the binding of the 
$J/\Psi$ meson is sensitive to the screening of the $c-\bar c$ potential  by a
quark-gluon plasma, and the meson bound state will not  survive in a hot enough 
quark-gluon plasma. Hence the original argument suggesting that a decrease of   the
observed
$J/\Psi$  yield could reveal the formation of the quark gluon plasma. 
An alternative scenario for the $J/\Psi$ suppression involves 
$J/\Psi$ collisions  with hard ``deconfined'' gluons present in the quark-gluon
plasma
\cite{Kharzeev94}. 

The first run of
experiments at CERN indeed showed that the rate of $J/\Psi$ production was
less than the rate expected from extrapolations of nucleon-nucleon collisions. But
it soon appeared that this phenomenon, as well as the corresponding one observed in
proton-nucleus collisions, could be accounted for  by what is usually referred to as
nuclear absorption
\cite{Capella88,GH88}.  A
$J/\Psi$   produced somewhere in the nucleus  has to cross a certain region of
nuclear matter before escaping, and because it can interact inelastically with
nucleons on its way out, it may be destroyed. A survival probability can then be
defined,
$\exp\{-L/\lambda\}$, where $L$ is the distance traveled by the $J/\Psi$ in
nuclear matter, and $\lambda=1/(n\sigma_{abs})$ an absorption mean free path  with 
 $n$ the nuclear density. Several analysis lead
to a value of
$\sigma_{abs}$ of the order of 6 to 7 mb \cite{Nardi97}. 

The fact that the Pb-Pb data \cite{Gonin96} do not obey this simple behaviour was a
surprise. And the temptation to speculate about a new mechanism at work has been
irresistible \cite{BO96,Kharzeev96}. Interestingly, if one assumes that the extra
suppression observed in Pb-Pb collisions is a local phenomenon (in space-time),
sensitive for instance only to the local energy density, one can account
quantitatively for the bulk of the data. But to prove that we are dealing with a
phenomenon  related indeed to local energy density one would need data at lower beam
energy, which is hard, if not impossible, to get. Alternatively, one could  explore
smaller systems for which the new mechanism would set in within the covered
$E_T$ range. 

While 
the present NA50 data  show clear deviations from the normal nuclear
absorption pattern, whether or not they present a threshold remains a
subject of controversy.  Observing a threshold behaviour
would be very suggestive of a qualitative change in the properties of matter at
some energy density. Progress has been made in the analysis of both  the low $E_T$
and the high $E_T$ regions, with in the latter case the elimination of
spurious reinteractions in the target. Worth emphasizing is the development of a new
method of analysis making use of minimum bias events; the new procedure  removes in
particular the statistical fluctuations in the Drell-Yan spectrum, which was a
limitation in previous analysis \cite{Cicalo99}.

\section{THE EARLY STAGES OF NUCLEAR COLLISIONS AT RHIC AND LHC}

With RHIC coming, one may attempt to extrapolate the knowledge gained at the AGS
and SPS to higher energies. The corresponding ``predictions'' of various models will
be reviewed at the end of this meeting. Let me concentrate here on a few conceptual
issues of relevance when exploring higher energies.

At high energy, semi-hard interactions leading to minijets are believed to play an
important role. They start to compete with ``soft'' phenomena presumably already at
RHIC where they may contribute a significant fraction of the total transverse
energy \cite{BM87,KLLEKL87}. The reasons why theorists are
interested in this regime is because it is one where one could hope to do reliable
calculations form  first principles in QCD.  The regime is indeed one of weakly
coupled many particles, which is perhaps amenable to a classical description. 

In fact Monte Carlo simulations, the so-called parton cascade calculations
\cite{Wang,KG} treat partons  as free
particles and study their evolution, taking into account QCD interactions, and
assuming that the initial distributions in  phase-space are given by the structure
functions of the nuclei.  These calculations provide a  detailed description, at the
partonic level, of the beginning of a nucleus-nucleus collision. They allow the study
of thermalisation, the build up of energy density and so on. However, they raise a
number of theoretical questions. To which extent  can partons be treated as
classical particles? What is the role of quantum mechanical coherence effects? 

In the last few years, efforts have been made to provide a more satisfying
theoretical framework. Simple  and
and physically appealing pictures of the initial wave function have been
constructed \cite{MCLRV94}.   But relating  these initial wave
functions to the initial conditions for parton cascades, that is understanding how
the gluon initially in the wave functions are freed in the collision, remains an open
problem
\cite{AHMueller99}.

\section{CONCLUSIONS}

The analysis of the various  data collected from BNL/AGS and
CERN/SPS have led to a truly impressive progress in our understanding of
nucleus-nucleus collisions at high energy:   A-A collisions  are
clearly distinct from N-N collisions. Large  energy densities are
produced, providing appropriate conditions for the creation of a
quark-gluon plasma. Collective behaviour of matter is seen and we have hints from 
several observables that  quark degrees of  freedom play
a role in the collision dynamics.  There are good indications that thermal and
chemical equilibrium may be  reached at freeze-out. 
The space-time pictures of the collisions are becoming more and more precise with flow
and interferometry measurements, and the large multiplicities of Pb-Pb collisions
allow for promising event by event analysis.  Flows patterns start to be used
to reveal details of the equation of state. 
 Finally, several anomalies  have been identified in observables considered as
potential ``signatures'' of the  quark-gluon plasma.

It is  fair to say, however, that we are not yet in a position  to offer to
outsiders to the field  compelling evidence that quark-gluon plasma has been
produced. For one thing, our theoretical picture of the quark-gluon plasma is still
very incomplete.  Calculating the properties of the
quark-gluon plasma, even in equilibrium, turns out to be a difficult task, beyond
the trivial level of the free gas (on which most experimental analyses rely). And the
off-equilibrium properties of the plasma are essentially unknown. Thus we have so far
no unique ``signature",  and what we are looking for is 
 evolving as we progress in our understanding of 
 nucleus-nucleus collisions. But the progress which has been made 
 since the very first run of experiments is really remarkable, and gives great
confidence in the future of the field. And with RHIC coming out now, there are all
reasons to be optimistic.

\end{document}